\documentstyle[12pt,a4wide]{article}
\def\PLB{Phys.~Lett.~B}
\def\NPB{Nucl.~Phys.~B}
\def\NP{Nucl.~Phys.}
\def\PRL{Phys.~Rev.~Lett.}

\def\PRP{Phys.~Rep.}
\def\PRD{Phys.~Rev.~D}
\def\MPL{Mod.~Phys.~Lett.}

\begin{document}

\begin{titlepage}
\begin{flushright}
UAB-FT-312
\end{flushright}
\vskip0.3cm
\begin{center}
\boldmath
{\large\bf SUPERSYMMETRIC CORRECTIONS TO \\
\ \\ MASSES AND WIDTHS}
\unboldmath
\vskip1cm
\newlength{\estrecho}
\setlength{\estrecho}{0.15\textwidth}
\addtolength{\textwidth}{-\estrecho}
{\sc P. Gosdzinsky}\\[1cm]
Grup de F\'\i sica Te\`orica,\\
Departament de F\'\i sica\\
and \\
Institut de F\'\i sica d'Altes Energies,\\[0.5cm]
Universitat Aut\`onoma de Barcelona,\\
E-08193 Bellaterra (Barcelona), SPAIN.\\[6cm]
\begin{abstract}
Supersymmetric radiative corrections to the partial width $\Gamma (\tau
\rightarrow e \nu \bar{\nu})$ and the masses
of the $W$ and $Z$ gauge bosons are calculated in different  renormalization
schemes. The importance of the non-universal
contributions is analyzed.
\end{abstract}
\end{center}
%
%
%
\addtolength{\textwidth}{\estrecho}
\end{titlepage}
\section{\bf INTRODUCTION}
The minimal supersymmetric standard model (MSSM) is one of the most
interesting extensions of the standard model of the
electroweak interactions (SM) \cite{SMOK}. Although the SM, a gauge theory
based on a spontaneously broken $SU(2)\times U(1)$
describes successfully all observed phenomena, it leaves several theoretical
questions
unanswered. Some of these problems, such as the hierarchy problem,
could be solved by supersymmetry (SUSY) \cite{HK,N}. Unfortunately, no
supersymmetric particle has been detected, and we do not
even know if SUSY has anything to do with the real world.  However, if SUSY
is realized in nature, it could be possible to detect
supersymmetric signals through radiative effects.

Several analysis of SUSY radiative corrections already exist in the
 literature (see, for example, \cite{BFGH,DHY,MDKH,PGS1,PGS2}
and references therein).
In this paper, we focus our attention on universal and full
one loop radiative corrections to the partial
width $\Gamma(\tau\rightarrow e \nu \bar\nu)$ and the masses of the $W$
and $Z$ gauge bosons in different
renormalization schemes. Sleptons and -inos (gauginos and higgsinos) will be
included in the analysis. The
phenomenological bounds
\begin{equation}
m_{\tilde l^+}\ge\ 45 \ {\rm GeV},\   m_{\tilde \nu}\ge 30\ {\rm GeV},\
M_{\Psi^+}\ge 45 \ {\rm GeV}
\end{equation}
will be respected.
In the -ino sector, we assume the GUT relation
\begin{equation}
M'={5\over 3} \tan^2(\theta) M.
\label{gutrel}
\end{equation}
The charged sleptons of each family are supposed to be degenerate in mass.
Their mass is related to
the sneutrino mass through
\begin{equation}
m^2_{\tilde l^+}=m^2_{\tilde \nu}+M_W^2 {\tan^2(\beta)-1 \over
\tan^2(\beta)+1},
\label{masal}
\end{equation}
where $\tan(\beta)$ is the ratio of the v.e.v. of the Higgs giving mass to
the up quarks and  the v.e.v. of the Higgs
giving mass to the down quarks.
We do not consider squark contributions, but they can easily be included:
Once their masses and
mixing angles are given, one just adds their (universal) contributions
$\delta M_W^{squarks}$ and $\delta M_Z^{squarks}$
to the given results. As we will see in sec. 4, $\Gamma (\tau \rightarrow e
\nu \bar{\nu})$ is not affected by
squark contributions.

This paper is organized as follows: In section 2, radiative corrections to
the $W$ mass are computed in the
popular $(\alpha, G_\mu, M_Z)$ scheme. In section 3, we consider radiative
corrections to the $Z$ mass in the
$(\alpha, G_\mu, M_W / M_Z\equiv \cos(\theta))$ scheme. Clearly, some steps
in these directions have already
been done (see, for example \cite{BFGH}). However, analyses including
 process dependent contributions (i.e.
vertex and box diagrams, as well as external wave function renormalization
effects),  which could be
important \cite{PGS1,PGS2} are, to my knowledge, still missing.
In section 4, we say a few words on $\tau$ decays, and finally,
section 5 will be devoted to the discussions and conclusions.

\section{RADIATIVE CORRECTIONS TO $M_W$}
Suppose we are given the fine structure constant $\alpha$, the Fermi
constant $G_\mu$ and the mass of the $Z$ boson $M_Z$.
Then, at tree-level, the following relation holds \cite{CH}:
\begin{equation}
{\bar M_W}^2= {M_Z^2 \over 2}\left(1+\sqrt{1-{4\pi\alpha \over \sqrt{2}
G_\mu M_Z^2}} \ \right).
\label{masawz}
\end{equation}
After inclusion of radiative corrections, eq.(\ref{masawz}) will be
modified, but it still holds for bare quantities.
Let us now introduce
\begin{equation}
{s_1}^2={1\over 2}\left(1-\sqrt{1-{4\pi\alpha \over \sqrt{2} G_\mu
M_Z^2}}\ \right) \ ,\qquad {c_1}^2+{s_1}^2=1,
\end{equation}
\begin{equation}
M_W^2={M_W^o}^2+\Pi_W(M_W^2)\equiv{M_W^o}^2(1+\delta_W),
\end{equation}
\begin{equation}
M_Z^2={M_Z^o}^2+\Pi_Z(M_Z^2)\equiv{M_Z^o}^2(1+\delta_Z),
\end{equation}
\begin{equation}
\alpha=\alpha^o(1+\delta_\alpha),
\end{equation}
\begin{equation}
G_\mu=G_{\mu}^o(1+\delta_\mu).
\label{gamanu}
\end{equation}
After substituting the bare quantities $M_W^o$, $M_Z^o$, $\alpha^o$ and
 $G_{\mu}^o$ in eq.(\ref{masawz}) and doing
some algebra, we find
\begin{equation}
M_W^2= M_Z^2 c_1^2\left\{1-{s_1^2\over c_1^2-s_1^2}(\delta_\mu
 -\delta_\alpha)-{c_1^2\over c_1^2-s_1^2}\delta_Z+\delta_W\right\}.
\end{equation}
 $\delta_{\alpha}^{SUSY}$, $\delta_{\mu}^{SUSY}$, $\delta_Z^{SUSY}$
and $\delta_Z^{SUSY}$ are given by
\begin{equation}
\delta_{\alpha}^{SUSY}=\left\{ {\Pi_{\gamma \gamma}(k^2)\over k^2}^{SUSY}
 \right\}
_{k^2\rightarrow 0},
\end{equation}
\begin{equation}
\delta_\mu^{SUSY}=-{\Pi_W(0)\over M_W^2}^{SUSY}+\delta_{\mu (SUSY)}^{NSE},
\end{equation}
\begin{equation}
\delta_W^{SUSY}={\Pi_W(M_W^2)\over M_W^2}^{SUSY},
\end{equation}
\begin{equation}
\delta_Z^{SUSY}={\Pi_Z(M_Z^2)\over M_Z^2}^{SUSY},
\end{equation}
where $\delta_{\mu (SUSY)}^{NSE}$ is a process-dependent contribution which
 depends on $m_{\tilde \nu _e}$,  $m_{\tilde \nu _\mu}$,
the selectron masses and mixing angles, the smuon masses and mixing angles,
 $\tan(\beta)$, and the -ino masses and mixing-matrices.
The formul\ae\ for  $\Pi_\gamma(k^2)$,
$\Pi_W(k^2)$, $\Pi_\gamma (k^2)$ and  $\delta_{\mu(SUSY)}^{NSE}$   can be
found in the literature \cite{GS1,GS2}
and will not be listed here.
The universal contributions to the mass shift $\delta M_W\equiv M_W-\bar M_W$
are given by
\begin{eqnarray}
& &{\displaystyle (\delta M_W^{SUSY})_{UNIV}={M_W\over 2}\left\{{s_1^2\over
c_1^2-s_1^2}
\left({\Pi_W(k^2)\over M_W^2}^{SUSY} +
{\Pi_{\gamma \gamma}(k^2)\over k^2}^{SUSY}\right)\right.}\nonumber\\
& &{\displaystyle \left.
  -{c_1^2\over c_1^2-s_1^2}{\Pi_Z(k^2)\over M_Z^2}^{SUSY}
+{\Pi_W(M_W^2)\over M_W^2}^{SUSY}\right\}_{k^2\rightarrow 0}},
\label{wun}
\end{eqnarray}
the non-universal contributions are given by
\begin{equation}
(\delta M_W^{SUSY})_{NON-UNIV}=-{M_W\over 2}{s_1^2\over c_1^2 -s_1^2}
\delta_{\mu(SUSY)}^{NSE},
\label{wnoun}
\end{equation}
and the full contributions read
\begin{equation}
{\displaystyle
\delta M_W^{SUSY}={M_W\over 2}\left\{-{s_1^2\over c_1^2-s_1^2}
\left(\delta_\mu^{SUSY}-\delta_\alpha^{SUSY}\right)
-{c_1^2\over c_1^2-s_1^2}\delta_Z^{SUSY}+\delta_W^{SUSY}\right\}.
}
\end{equation}

In fig.1, we show the universal (fig.1.a) and full (fig.1.b) contributions
to $\delta M_W$ for $\tan (\beta)=2$ and
$m_{\tilde \nu _e}=m_{\tilde \nu _\mu}=m_{\tilde \nu _\tau}=$ 50 GeV.
The masses of the charged sleptons are given in
eq.(\ref{masal}). The blank regions are phenomenologically excluded by the
mass limit $ M_{\Psi^+}\ge$ 45 GeV.
comparison of fig.1.a and fig.1.b shows that the non-universal contributions
turn out to be small.

In the main range of the allowed $(\mu,M)$ space the full corrections
(fig.1.b) lie between $\sim$60 and $\sim$120 MeV,
mainly due to the universal contributions (eq.(\ref{wun})), which are
about one order of
magnitude bigger than the non-universal contributions (eq.(\ref{wnoun})).
Similar analyses, with heavier sleptons, show that the mass shift decreases
as the sfermions become
heavier. For example, for $m_{\tilde \nu _e}=m_{\tilde \nu _\mu}=m_{\tilde
\nu _\tau}=$ 200 GeV and $\tan(\beta)=2$,
the full corrections lie
between $\sim -10$ and $\sim 50$ MeV (in the main range of the allowed
$(\mu,M)$ space).
In this case, the non-universal contributions turn out to be very small.
They are not bigger than 20 MeV
in the whole allowed space, and not bigger than 10 MeV in a wide region.
These mass shifts lie bellow the
present experimental uncertainty $\Delta M_W=$ 250 MeV . However, $M_W$
will be measured with an accuracy of
$\sim 100$ MeV
in the near future and thus, the experimental uncertainty and SUSY mass
shifts will become comparable,
provided the sparticles are light enough and squarks have been included
in the analysis.

\section{RADIATIVE CORRECTIONS TO $M_Z$}
Suppose now that we are given the fine structure constant $\alpha$, the
Fermi constant $G_\mu$ and the ratio of the
masses of the $W$ and the $Z$ bosons, $M_W / M_Z \equiv \cos(\theta)$.
Then, at tree-level, the following relation holds:
\begin{equation}
{\overline M}_Z^2={\pi\alpha\over \sqrt{2} G_\mu \sin^2(\theta)
\cos^2(\theta)}
\label{masazc}
\end{equation}

As in the previous section, eq.(\ref{masazc}) will get modified when
radiative corrections are included, but it
still holds for bare quantities.
Substituting bare quantities in eq.(\ref{masazc}), we find (see also
\cite{CPM})
\begin{equation}
M_Z^2={\pi\alpha\over \sqrt{2} G_\mu \sin^2(\theta) \cos^2(\theta)}\left\{1
-\delta_\alpha+\delta_\mu+{2s_2^2-1\over s_2^2}\delta_W
+{1-s_2^2\over s_2^2}\delta_Z\right\},
\end{equation}
where $s_2=\sin(\theta)$. The universal contributions to $\delta M_Z \equiv
M_Z-\overline M_Z$ are given by
\begin{eqnarray}
&&{\displaystyle (\delta M_Z^{SUSY})_{UNIV}={M_Z\over 2}
\left\{-{\Pi_{\gamma \gamma}(k^2)\over k^2}^{SUSY}-{\Pi_W(k^2)\over
M_W^2}^{SUSY}\right.
}\nonumber
\\
&&\displaystyle{\left.
+{2s_2^2-1\over s_2^2}{\Pi_W(M_W^2)\over M_W^2}^{SUSY}+{1-s_2^2\over s_2^2}
{\Pi_Z(M_Z^2)\over M_Z^2}^{SUSY}\right\}_{k^2\rightarrow 0}
},
\label{zun}
\end{eqnarray}
the  non-universal contributions are given by
\begin{equation}
(\delta M_Z^{SUSY})_{NON-UNIV}={M_Z\over 2}\delta_{\mu(SUSY)}^{NSE},
\label{znoun}
\end{equation}
while the full contributions read
\begin{equation}
\delta M_Z^{SUSY}={M_Z\over 2} \left\{ -\delta_\alpha^{SUSY}+
\delta_\mu^{SUSY}+{2s_2^2-1\over s_2^2}\delta_W^{SUSY}+
{1-s_2^2\over s_2^2}\delta_Z^{SUSY} \right\}.
\end{equation}

In fig.2, $(\delta M_Z^{SUSY})_{UNIV}$ (fig.2.a) and
$(\delta M_Z^{SUSY})_{FULL}$ (fig.2.b) are displayed for $\tan (\beta)=2$
and $m_{\tilde \nu_e}=m_{\tilde \nu_\mu}=m_{\tilde \nu_\tau}$=50 GeV.
The masses of the charged sleptons are given
in eq.(\ref{masal}). The blank regions are phenomenologically excluded by
the mass limit $ M_{\Psi^+}\ge$ 45 GeV.
Comparison of fig.2.a and fig.2.b shows that, the non-universal
contributions turn out to be small. As in the previous
section, the universal contributions  (eq.(\ref{zun})) are about one
order of magnitude bigger the non-universal
contributions (eq.(\ref{znoun})).

The full mass shifts (fig.2.b) lie between $\sim -150$ and $\sim -310$ MeV
in the main region of the allowed $(\mu,M)$ space.
Similar analyses show that these mass-shifts decrease for heavier
sleptons, e.g., for
$m_{\tilde \nu_e}=m_{\tilde \nu_\mu}=m_{\tilde \nu_\tau}$=200 GeV and
$\tan(\beta)=2$, the radiative corrections
lie between $\sim 30$ and $\sim-110$
MeV (in the main range of the allowed $(\mu,M)$ space). As in the
previous section, the non-universal contributions are found to be small.
For light sleptons,  these mass shifts are bigger than the present
experimental error $\Delta M_Z=20$ MeV \cite{PPDB}! Unfortunately, at
tree-level $M_Z$ can only be predicted with a precision
of $\sim$
670 MeV. This tree-level error is due to the present experimental
uncertainty $\Delta M_W =$ 250 MeV. However,
for light sparticles, SUSY mass-shifts and the tree-level uncertainty
will become comparable in the future, once $M_W$
is measured with an accuracy of $\sim$ 100 MeV and squarks have been
included in the analyses.

\section {$\tau$ decays}
Let us now say a few words on $\tau$ decays. The $\tau$ lifetime,
$\tau_\tau$, which can be derived from direct measurements
of the impact parameter, can also be calculated from the measured branching
ratio
\begin{equation}
B(\tau\rightarrow e \nu \bar\nu ))={\Gamma(\tau\rightarrow e \nu \bar\nu
 )\over \Gamma(\tau\rightarrow all)},
\end{equation}
where $\Gamma(\tau\rightarrow e \nu \bar\nu )$ (electroweak radiative
corrections are included, \cite{WMAS}) is given by
\begin{equation}
\Gamma(\tau\rightarrow e\nu \bar \nu)_\mu={G_\mu^2 m_\tau^5 \over
192 \pi^3}f({m_e^2\over m_\tau^2})
\left(1+{3\over 5}{m_\tau^2 \over M_W^2}\right) \left(1+{\alpha(m_\tau)\over
2\pi}({25\over 4}-\pi^2)\right),
\label{gamamutau}
\end{equation}
and $f(x)=1-8x+8x^3-x^4 -12x^2 \log(x)$.
Since both results for $\tau_\tau$ should agree, it is expected that
\begin{equation}
R_\tau\equiv{\Gamma(\tau\rightarrow e\nu \bar \nu )_\mu \over
B(\tau\rightarrow e\nu \bar \nu )
\Gamma(\tau\rightarrow all)}=1
\label{rtau}
\end{equation}
holds.
However, eq.(\ref{gamamutau}) may be affected by non-electroweak
corrections. Supersymmetric corrections to eq.(\ref{gamamutau}) can
be included by introducing a new Fermi constant $G_\tau$,
\begin{equation}
G_\tau=G_\mu^o(1+\delta_\tau^{SUSY}),
\label{gtau}
\end{equation}
where $\delta_\tau^{SUSY}=\delta_\mu^{SUSY}(\mu\leftrightarrow \tau)$.
Combining eq.(\ref{gamanu}) and eq.(\ref{gtau}), we find
\begin{equation}
G_\tau=G_\mu(1+\delta_\tau^{SUSY}-\delta_\mu^{SUSY}).
\label{gtauren}
\end{equation}
Thus, if supersymmetric corrections are included, the partial
width $\Gamma(\tau\rightarrow e\nu \bar \nu)$
reads
\begin{equation}
\Gamma(\tau\rightarrow e\nu \bar \nu)={G_\tau^2 m_\tau^5 \over 192
\pi^3}f({m_e^2\over m_\tau^2})
\left(1+{3\over 5}{m_\tau^2
\over M_W^2}\right) \left(1+{\alpha(m_\tau)\over 2\pi}
({25\over 4}-\pi^2)\right),
\label{gammaok}
\end{equation}
and $R_\tau$ (eq.(\ref{rtau})) will deviate from 1. Supersymmetric
corrections to $R_\tau$, i.e.,
\begin{equation}
\delta R_\tau^{SUSY}=2(\delta_{\mu (SUSY)}^{NSE}-\delta_{\tau (SUSY)}^{NSE}),
\label{drsusy}
\end{equation}
where
$\delta_{\tau (SUSY)}^{NSE}$ is a process-dependent contribution which
depends on $m_{\tilde \nu_e}$,  $m_{\tilde \nu _\tau}$,
the selectron masses and mixing angles, the stau masses and mixing
angles, $\tan(\beta)$, and the -ino masses and mixing-matrices,
will arise if $m_{\tilde \nu_\tau}\neq m_{\tilde\nu_\mu}$
(and $m_{\tilde \tau}\neq m_{ \tilde \mu}$, eq(\ref{masal})). Thus,
$R_\tau$ is sensitive to mass splittings between the $\tilde \nu_\mu$
and the $\tilde \nu_\tau$
(and between the smuon and the stau).
Notice that there are no universal contributions to $\delta R_\tau^{SUSY}$.
Furthermore,  as already pointed out in the introduction, $R_\tau$ is
not affected by squark contributions.
Combining eqs.(\ref{gamamutau}), (\ref{gtauren}), (\ref{gammaok})
and (\ref{drsusy}), we find
\begin{equation}
\Gamma(\tau \rightarrow e \nu \bar \nu)=\Gamma(\tau \rightarrow e
\nu \bar \nu)_{\mu}(1-\delta R_\tau^{SUSY})
\end{equation}
and
\begin{equation}
\delta \Gamma(\tau \rightarrow e \nu \bar \nu)^{SUSY}=-\Gamma(\tau
\rightarrow e \nu \bar \nu)\delta R_\tau^{SUSY}.
\end{equation}

Substituting the updated (and preliminary) data \cite{MD}
\begin{equation}
m_\tau=(1777.1 \pm 0.5) {\rm MeV},
\label{newmt}
\end{equation}
\begin{equation}
\tau_\tau=(295.7 \pm 3.2) {\rm fs},
\label{newtt}
\end{equation}
\begin{equation}
B_e=(17.76 \pm 0.15 )\%,
\label{newbt}
\end{equation}
in eq.(\ref{rtau}) and eq.(\ref{gammaok}), we find
\begin{equation}
R_\tau=1.020 \pm 0.014.
\label{numrtau}
\end{equation}
Let us now turn to the numerical analysis. In fig.3,
$\delta_{\tilde l(SUSY)}^{NSE}$, $(\tilde l= \tilde \mu, \tilde \tau)$ is
displayed for $\mu=-50$ GeV, $M=$ 100 GeV and $\tan (\beta)=2$. The
masses of the charged sleptons are given in eq.(\ref{masal}).
By combining eq.(\ref{drsusy}) and the results in fig.3, we easily
obtain the value of $\delta R_\tau$, which turns out
to be small $(\vert  \delta R_\tau \vert \leq 0.25 \%$).
Similar analyses show that $\vert \delta_{\tilde l (SUSY)}^{NSE} \vert$
(and thus, $\vert \delta R_\tau ^{SUSY}\vert$ and
$\vert \delta \Gamma(\tau\rightarrow e \nu \bar \nu)^{SUSY} \vert$)
decrease as $\vert \mu \vert$ and $\vert M \vert$ increase.

I would not like to conclude this section without saying a few words on
the $\tau$-decay puzzle. For the averages
$m_\tau=1784.1_{-3.6}^{+2.7}$ MeV and $\tau_\tau=(0.305\pm 0.006)
10^{-12}$ sec, \cite{PPDB} the experimental branching ratios
for $\tau$ decays where found to be smaller than the theoretical
predictions by a factor $0.95\pm 0.02$. Several
attempts have been made to solve this puzzle with help of new physics
 \cite{WJS,WHTS,GHK}. Supersymmetric contributions
to $\Gamma(\tau\rightarrow e \nu \bar\nu)$, and thus, to
$B(\tau\rightarrow e \nu \bar\nu)$
are too small to solve the $\tau$ puzzle.
However, the most recent measurements of the $\tau$ mass and lifetime
(eqs.(\ref{newmt}) and (\ref{newtt}), see also ref.\cite{MD}) seem
to resolve these discrepancies.

\section{CONCLUSIONS}

Supersymmetric one loop corrections to the partial width
$\Gamma(\tau\rightarrow e \nu \bar\nu)$ and the masses of
the $W$ and $Z$ gauge bosons have been calculated. All the
contributions (i.e. universal and non-universal) have been taken
into account. $\Gamma(\tau\rightarrow e \nu \bar\nu)$ is only affected
by non-universal contributions, which turn out
to be very small.

The mass shifts $\delta M_W^{SUSY}$  ($\delta M_Z^{SUSY}$) where found to
lie bellow the present
experimental (tree-level) errors, but in the future they might become
comparable to these uncertainties, once $M_W$ is measured with a precision
of $\sim$ 100 MeV.
When the top quark mass will be known and
squarks have been included in the analyses, SUSY might produce visible
signals through both, $M_W$ and $M_Z$ mass shifts,
which go in opposite directions. Thus, a deviation of the standard model
prediction for $M_W$ (in the ($\alpha, G_\mu, M_Z$)
scheme) from
the  measured value $M_W$ {\em and} a deviation, {\em in opposite direction},
of the standard model prediction for $M_Z$
(in the $(\alpha, G_\mu, M_W/M_Z)$ scheme) from
the measured value $M_Z$  could be an indication for new physics
(not necessarily
supersymmetry), even if the deviations are not big. The sign and size
of these deviations would contain valuable information.

For light sparticles, the computed mass shifts are dominated by universal
contributions, which, are about one order
of magnitude bigger than the non-universal contributions (we have found
a small region, easily explorable at LEP II, where the
non-universal contributions are about 30 $\%$ of the universal
contributions), while for
heavier sparticles, both, the universal and the non-universal
contributions turn out to be small
(I have checked that this also holds for $\tan (\beta) \neq 2$).
The fact that $(\delta M_W^{SUSY})_{NON-UNIV}$
($(\delta M_Z^{SUSY})_{NON-UNIV}$) is not big can also be seen by combining
eq.(\ref{wnoun}) (eq.(\ref{znoun})) and the results displayed in fig.3.

{\bf Acknowledgments:} I acknowledge gratefully a grant from the
Generalitat de Catalunya.
 I am also grateful to S. Peris for the careful reading of the manuscript.

{\bf Figure Captions}

{\bf fig.1.}   Contour plots of $\delta M_W^{SUSY}$ (MeV) for
$m_{\tilde \nu}$= 50 GeV and $\tan (\beta)$=2.
In fig.1.a only universal contributions are displayed, while in fig.1.b
all contributions have been taken into account.
The blank regions are phenomenologically excluded by $M_{\Psi^+}\ge$ 45 GeV.

{\bf fig.2.}   Contour plots of $\delta M_Z^{SUSY}$ (MeV) for
$m_{\tilde \nu}$= 50 GeV and $\tan (\beta)$=2.
In fig.2.a only universal contributions are displayed, while in fig.2.b
all contributions have been taken into account.
The blank regions are phenomenologically excluded by $M_{\Psi^+}\ge$ 45 GeV.

{\bf fig.3.}  Contour plots for $\delta_{\tilde l (SUSY)}^{NSE}
(\%)$ for $\mu=-50$ GeV , $M=$ 100 GeV and $\tan(\beta)$=2.

\end{document}